\documentclass[11pt,a4paper]{article}
\pdfoutput=1
\usepackage{jheppub}
\usepackage{amssymb}
\usepackage{amsfonts}
\usepackage{graphicx}
\usepackage{epstopdf}
\usepackage{dcolumn}
\usepackage{amsmath}
\usepackage{latexsym,bm}
\usepackage{amsthm}
\usepackage{slashed}
\usepackage{float}
\usepackage{color}
\usepackage{url}

\def \be {\begin{equation}}
\def \ee {\end{equation}}
\def \bsp {\begin{split}}
\def \esp {\end{split}}
\def \bea {\begin{eqnarray}}
\def \eea {\end{eqnarray}}

\def\mc{\mathcal}

\def\P{\mathbb{P}}

\def\Z{\mathbb{Z}}
\def\F{\mathbb{F}}

\def\MM{{\cal  M}_{\rm max}}
\def\bmx{{B}_{\rm max}}

\title{The F-theory geometry with most flux vacua}

\author[a]{Washington Taylor,}
\author[a]{Yi-Nan Wang}

\affiliation[a]{Center for Theoretical Physics,\\
Department of Physics\\Massachusetts Institute of Technology\\
77 Massachusetts Avenue\\Cambridge, MA 02139, USA} 

\emailAdd{wati@mit.edu}
\emailAdd{wangyn@mit.edu}

\preprint{\today \hspace*{0.1in} MIT-CTP-4732}

\abstract{Applying the Ashok-Denef-Douglas estimation method to
  elliptic Calabi-Yau fourfolds suggests that a single elliptic
  fourfold ${\cal M}_{\rm max}$ gives rise to ${\cal O}
  (10^{272,000})$ F-theory flux vacua, and that the sum total of
the numbers of flux
  vacua from all other F-theory geometries is suppressed by a
relative
  factor of ${\cal O} (10^{-3000})$.  The fourfold $\MM$ arises from a
  generic elliptic fibration over a specific toric threefold base
  $B_{\rm max}$, and gives a geometrically non-Higgsable gauge group
  of $E_8^9 \times F_4^8 \times (G_2 \times SU(2))^{16}$, of which we
  expect some factors to be broken by G-flux to smaller groups.  
It is not possible to tune an $SU(5)$ GUT group on any further
divisors in $\MM$, or even an $SU(2)$ or $SU(3)$, so the standard
model gauge group appears to arise in this context only from a broken
$E_8$ factor.
The results of this paper can either be interpreted as providing a framework for
  predicting how the standard model arises most naturally in F-theory
  and the types of dark matter to be found in a typical F-theory
  compactification, or as a challenge to string theorists to explain
  why other choices of vacua are not exponentially unlikely compared
  to F-theory compactifications on $\MM$.}

\keywords{}

\begin{document}

\maketitle

\flushbottom

\section{Introduction}

The apparent existence of an enormous number of possible consistent 4d
vacuum solutions to string theory poses practical and philosophical
challenges for the predictive power of the theory.  On one hand,
Weinberg's argument \cite{Weinberg} and the possibility of
cosmological inflation and vacuum tunneling in a multiverse fit
naturally with the many vacua of string theory into an anthropic
explanation for the observation of a small but nonzero cosmological
constant, roughly $10^{-120}$ in natural units \cite{cc}.  On the
other hand, there is as yet no sound methodology for computing the
relative abundance of different string vacuum solutions, and we are far from
a complete understanding of the full set of possible vacuum
solutions with supersymmetry, let alone of those without
supersymmetry.

The largest numbers of different string vacua studied
to date arise in the form of ``flux compactifications''
\cite{Grana:2005jc,
Douglas:2006es,
Denef-F-theory}.  A flux compactification is a string
compactification on a geometric space ${\cal M}$, combined with a
choice of generalized $p$-form fluxes, analogous to magnetic flux,
that thread various topological cycles on ${\cal M}$.  In general, the
set of fluxes is constrained by a tadpole condition ({\it e.g.},
from varying one
of the fields in the Lagrangian in a supergravity approximation) so
that the number of flux vacua that can arise for any given geometry
becomes bounded, 
though it can be exponentially large.  For type IIB string theory, the
number of flux compactifications on certain geometries is famously
estimated at $\sim {\cal O} (10^{500})$.  For type IIA string theory
there may be infinite families of flux vacua \cite{dgkt}, though there
are believed to be only a finite number of possibilities at any given
compactification scale \cite{Acharya:2006zw}.  The largest
concrete numbers of
flux vacua known arise in F-theory \cite{Vafa-F-theory, Morrison-Vafa-I,
  Morrison-Vafa-II}, a nonperturbative version of type IIB theory with
an axiodilaton that is allowed to vary over the compact manifold,
which is a complex threefold (real 6-manifold) in the case of
compactification to 4d.

In this paper we argue that virtually all F-theory flux
compactifications may arise from a single geometry, ${\cal M}_{\rm
  max}$, and that all other F-theory flux compactifications taken
together may represent a fraction of $\sim {\cal O} (10^{-3000})$ of
the total set.  The geometry $\MM$, which was first identified in
\cite{Candelas-pr, lsw-4D}, is an elliptically fibered Calabi-Yau fourfold with
Hodge numbers $h^{1, 1} = 252, h^{3, 1} = 303,148$.  $\MM$ can be
understood as a generic elliptic fibration ({\it i.e.}, arising from a
generic Weierstrass model describing the axiodilaton) over a
particular base $B_{\rm max}$, which is a complex threefold with a
simple description in toric geometry.  The identification of the base
$\bmx$ gives a great deal of information about the possible forms of
the low-energy 4d physics that can arise from an F-theory flux
compactification on $\bmx$.  In particular, the geometric structure of
4d ``non-Higgsable clusters'' \cite{clusters, Anderson-WT, ghst,
4D-NHC, Halverson-WT} on $\bmx$ shows that in the absence of
fluxes and an
associated superpotential, the gauge group of the low-energy 4d theory
at a generic point in the moduli space of $\MM$ would be
\begin{equation}
 G_{\rm max} = E_8^9 \times F_4^8 \times (G_2 \times SU(2))^{16} \,,
\end{equation}
with charged matter transforming in the bifundamental
representation of each $G_2 \times SU(2)$ factor.  The presence of flux (G-flux) can, and
likely will in most flux vacua, break some or all of these factors
into subgroups
in a fashion amenable to statistical analysis on the large set of possible fluxes.

Thus, it seems natural to speculate that the multiverse could be
dominated by F-theory flux vacua on $\MM$, among which a typical
vacuum would have as gauge group a subgroup of $G_{\rm max}$.  The
standard model could be realized in this scenario through breaking of
an $E_8$ sector to $SU(3) \times SU(2) \times U(1)$, with the
remaining factors (or subgroups thereof) of $G_{\rm max}$ providing a
set of hidden dark matter sectors, connecting to the standard model
only through gravitational and nonperturbative physics.  If this
story, or some modified version thereof, is correct, it provides a
predictive framework for string theory despite the plethora of vacuum
solutions.

In \S\ref{sec:MM} we describe the geometry of $\MM$ in more detail.
In \S\ref{sec:physics}, we describe the possible physics of a typical
F-theory on $\MM$ vacuum, and discuss the relative numbers of vacua
associated with other compactification geometries.  In
\S\ref{sec:issues}, we describe some of the many possible ways in
which this nice story may break down.

\section{F-theory on the fourfold $\MM$}
\label{sec:MM}

In this section we give a description of the fourfold $\MM$ and the
corresponding F-theory models.  The notation used follows that of
\cite{MC}, where a more detailed description is given of toric
F-theory compactifications of this type.
For more general background on F-theory and compactifications,
see \cite{Morrison-TASI, Denef-F-theory, WT-TASI}.

\subsection{The geometry of
$\MM$ as an elliptic fibration}

The complex four-manifold $\MM$ is an elliptically fibered Calabi-Yau
manifold with Hodge numbers and fourth Betti number
\begin{eqnarray}
h^{1, 1} & = &  252\\
h^{2, 1} & = &  0\\
h^{3, 1} & = & 303,148 \label{eq:h31}\\
h^{2, 2} & = & 4 (h^{1, 1}+h^{3, 1}) + 44-2h^{2, 1} = 1,213,644\\
b_4 & = & 2+2h^{3,1}+h^{2,2}=1,819,942\,.
\end{eqnarray}
The Euler character is thus
$\chi = 6 (8 +h^{1, 1} +h^{3, 1}-h^{2, 1})
= 1,820,448$.  This geometry was identified in
\cite{Candelas-pr, lsw-4D} using toric and Landau-Ginzburg model
approaches to constructing Calabi-Yau fourfolds.  
The four-manifold $\MM$ is the Calabi-Yau fourfold with the largest
known value of $h^{3,1}$.  This fourfold and its mirror share the
largest known Euler character, and the potential importance of these
manifolds for F-theory was commented on  in
\cite{Douglas-essay}.

The geometry of $\MM$ can be understood as that of a generic elliptic
fibration over a complex threefold base $\bmx$.
$\bmx$ is itself formed as a $B_2$ bundle over $\P^1$, where $B_2$ is
a toric surface characterized by a closed cycle of toric divisors
(curves, corresponding to rays in the toric fan) with
self-intersections $0, + 6,$
-12//-11//-12//-12//-12//-12//-12//-12//-12, where $//$ denotes the
sequence of self-intersections $-1, -2, -2, -3, -1, -5, -1, -3, -2,
-2, -1$.  $B_2$ itself supports a generic elliptic Calabi-Yau
threefold that has Hodge numbers $(251, 251)$ \cite{mt-toric, Hodge}.
In the language of toric geometry \cite{Fulton, Danilov}, $B_2$ is described by a 2D fan
generated by a set of rays $v_i \in N =\Z^2$.  The rays $v_i$ can be
taken to be 
\begin{eqnarray}
v_1 & = &  (-1, -12)\\
v_2 & = &  (0, 1)\\
v_3 & = & (1, 6)\\
\vdots & & \vdots\\
v_{99} & = & (0, -1) \,.
\end{eqnarray}
The intermediate rays can be determined by the condition
$v_{i -1} + v_{i + 1}+ (C_i \cdot C_i) v_i = 0$, where $C_i \cdot C_i$
is the self-intersection of the $i$th curve.
From the rays $v_i$ we can construct the toric fan for $\bmx$, which
is given by the rays
\begin{eqnarray}
w_0 & = & (0, 0, 1)\\
w_i & = & (v_i, 0), \;\;\;\;\; 1 \leq i \leq 99\\
w_{100} & = & (84, 492, -1) = (12v_{15}, -1) \,,
\end{eqnarray}
where $C_{15}$ is the curve in $B_2$ of self-intersection $-11$.  The
3D cones of the fan for $\bmx$ are spanned by $(w_0, w_i, w_{i + 1})$
and $(w_{100}, w_i, w_{i + 1})$, including the cyclic case
$(w_0/w_{100}, w_{99}, w_1)$.  This manifestly gives $\bmx$ the
structure of a $B_2$ bundle over $\P^1$, where the toric projection
onto the third axis corresponds to the fibration structure.  The
``twist'' in this bundle is characterized by the offset $12v_{15}$ of
the ray $w_{100}$.  The 3D polytope containing the vertices $w_i$ is
defined by the tetrahedron spanned by $w_0, w_1, w_2, w_{100}$.

A direct toric computation of the number of monomials in the generic
Weierstrass model $y^2 = x^3 + fx + g$ along the lines of
\cite{Anderson-WT, Halverson-WT, Hodge-btw}, with an
appropriate offset for automorphisms, reproduces the
value (\ref{eq:h31}).  Alternatively, constructing the generic
elliptic fibration over $\bmx$ as a hypersurface in a toric variety by
extending the polytope for $\bmx$ to the polytope of a $ \P^{2, 3, 1}$
fibration over $\bmx$ gives a construction that is linearly equivalent
to that described in \cite{Candelas-Skarke}. For the details of the linear transformation see Appendix \ref{app:linear}.

\subsection{Geometric non-Higgsable structures on $\MM$}

A useful tool in classifying geometries of base manifolds that support
elliptic fibrations for F-theory is the ``non-Higgsable'' geometric
structure of the base.  Non-Higgsable clusters refer to gauge groups
or products of gauge groups that arise from connected sets of codimension one Kodaira
singularities in generic elliptic fibrations over a given base.
Non-Higgsable clusters for base surfaces were classified in
\cite{clusters} and have provided a valuable tool in classifying
elliptic threefolds and associated 6d supergravity theories
\cite{mt-toric, Hodge, Martini-WT, Johnson-WT,  non-toric} as well as
6d superconformal field theories \cite{SCFT-1, SCFT-2, SCFT-3}.
Non-Higgsable clusters for base threefolds were systematically
analyzed in \cite{4D-NHC} and have been used in analyzing 4d F-theory
models in \cite{Anderson-WT, ghst, Halverson-WT, MC}.  For the base $\bmx$
the analysis of monomials in $f, g$ and the associated discriminant
$\Delta = 4f^3 + 27g^2$ in the Weierstrass model shows that there are
non-Higgsable gauge factors on precisely the divisors associated with
curves in $B_2$ that carry non-Higgsable gauge factors in the
corresponding 6d theory, where the $-12$ (and $-11$) curves carry
$E_8$ factors, the $-5$ curves carry $F_4$ factors, and the $-3, -2,
-2$ sequences each carry $G_2 \times SU(2)$ products with
bifundamental matter.  Thus, the geometrically non-Higgsable gauge
group of a generic elliptic fibration over $\bmx$ is
\begin{equation}
G_{\rm max} =
E_8^9 \times F_4^8 \times (G_2 \times SU(2))^{16} \,.
\label{eq:group}
\end{equation}
This group was originally associated with the elliptic Calabi-Yau
fourfold $\MM$ in \cite{Candelas-pr} using the method of
``tops,'' which describe both Higgsable and non-Higgsable gauge group factors.

While for F-theory compactifications to six dimensions, there is a
precise correspondence between geometrically non-Higgsable structure
and massless gauge groups and matter in the low-energy theory, the
connection is less transparent for 4d compactifications.  In
particular, as discussed in \cite{4D-NHC, Anderson-WT, ghst}, flux on
seven-brane world-volumes can break geometrically non-Higgsable gauge
factors, and G-flux produces a superpotential that can in principle
drive the theory to loci with a further enhanced gauge symmetry.  In
the case of $\bmx$, however, there is no possibility of acquiring further gauge symmetry
without producing codimension two singularities where $(f, g, \Delta)$
vanish to orders $(4, 6, 12)$, so it seems impossible to produce
additional nonabelian gauge groups (such as a GUT SU(5) or standard
model SU(3)$ \times $SU(2)) by tuning Weierstrass moduli, either by
hand or through forcing by a superpotential, without changing the base
or a hitting a superconformal point.
(For more detailed discussion see Appendix \ref{app:tuning}.) Thus, it seems that the only
gauge groups that can be realized on seven-branes
in flux vacua for F-theory
compactifications on $\MM$ will be subgroups of $G_{\rm max}$.  The
factors in $G_{\rm max}$  can generically be broken by nonzero G-flux
components, however.  An example of this was given in
\cite{Anderson-WT}, where flux on the world-volume of a set of
7-branes on a divisor carrying a
geometrically non-Higgsable $E_8$ gauge factor breaks the $E_8$
to $E_7$ in a setup with a clear heterotic dual geometry.  More
generally, we expect that $E_8$ factors can be broken to smaller
groups.  This seems like the most likely scenario for realizing the
standard model through F-theory on $\MM$, by a G-flux breaking of
$E_8$ to SU(3) $\times$ SU(2) $\times$ U(1), though other scenarios may
be possible.  

\subsection{The standard model and dark matter on $\MM$}

We do not attempt a thorough analysis here but make some comments on
how the standard model may emerge and the likely nature of dark matter
for F-theory compactifications on $\MM$.

Breaking $E_8$ to the standard model gauge group SU(3) $\times$ SU(2) $\times$ U(1) through fluxes  may occur in a variety of ways.
Some exploration of the possibility of realizing the standard model
gauge group by breaking an $E_8$ 
without using Wilson lines was described in \cite{Anderson-breaking}.
It would clearly be of interest to analyze such G-flux
induced breakings more systematically in this context.
$E_8$ can also be broken to the standard model group by first breaking to a unification group such as SU(5) or SO(10) and then breaking further through fluxes to SU(3)  $\times$ SU(2) $\times$ U(1), potentially connecting with the methods used in F-theory GUT models \cite{bhv, bhv2, Donagi-Breaking}.

As mentioned above, the details of how the standard model can arise
from a flux breaking of $E_8$ in an F-theory construction are not
fully understood.  It seems, however, that even if such a construction
is relatively unlikely compared to other breakings, this will impose
only a small penalty on the overall weighting of the number of flux
vacua.  In particular, for the case of $\MM$ a flux configuration is
characterized by $\chi/24\approx 75,000$ units of flux distributed among 
$b_4\approx 1,800,000$ possible cycles.  Thus, a typical individual cycle will be
given one unit of flux roughly 1/24 of the time, two units of flux
roughly 1/576 of the time, {\it etc.}. 
Because the number of configurations is so large, a simple statistical
model may suffice for estimating the fraction of models on $\MM$ that
have a given breaking pattern for $G_{\rm max}$

As discussed in \cite{Grimm-hkk, 
bkl-fluxes, Braun-Watari1,Braun-Watari2}, the space of the
4-form flux can be divided into horizontal, vertical and the remaining
components:
\be
H^4=H^4_{H*}\oplus H^{2,2}_{V*}\oplus H^{2,2}_{RM*}.
\ee
The horizontal component of the flux does not break any gauge
symmetry, and the rest of the components may break the gauge symmetry. For
the Calabi-Yau fourfold in the regime of $h^{1,1}\ll h^{3,1}$, the
size of the spaces $H^{2,2}_{V*}\oplus H^{2,2}_{RM*}$ are on the order of
$h^{1,1}$, hence they are negligible compared to $H^4_{H*}$ (as found,
for
example, for the elliptically fibered Calabi-Yau fourfolds mentioned in
\cite{Braun-Watari2}).

An important class of 4-cycles relevant for breaking a given $E_8$
factor are those that carry the ``Cartan flux'' $G=F\wedge \nu\in
H^{2,2}$\cite{bhv2,Donagi-Breaking,Braun-cycle}. Here $F\in H^{1,1}$
is the 2-form dual to the divisor carrying $E_8$, and $\nu\in H^{1,1}$
is the 2-form dual to the exceptional divisor in a resolution of this
Calabi-Yau fourfold.

For a given gauge group, the number of exceptional divisors is equal to its rank. If, for example a given breaking
pattern of the $E_8$ requires that four specific cycles get units of flux,
breaking the rank to 4, this is suppressed by a
factor\footnote{Considering
that there are 9 $E_8$ gauge groups, this possibility is really of
order $10^{-5}$} of order $10^6$. Obviously, further specific choices
may be needed to get 
exactly the standard model physics, but the tuning involved is
unlikely to approach the level of one part in $10^{3000}$, which would
be necessary for flux vacua from any other base to become relevant in
the standard statistical picture.

Regarding dark matter, due to the diffuse nature of the fluxes, it
seems that many of the additional factors in the gauge group
(\ref{eq:group}) will either be unbroken, or broken only slightly by
fluxes.  So we would expect that this model predicts about 30
independent decoupled dark sectors, each with its own nonabelian gauge
group of $E_8, F_4,$ $G_2 \times SU(2)$ or subgroup thereof.  Thus, in
this scenario most dark matter would not be directly coupled to the
standard model.  While this might help explain the absence so far of a
clear dark matter signal, it means that rather indirect means must be
used to get a quantitative test of the validity of this hypothesis.  A
detailed analysis of the potential behavior of decoupled dark matter
sectors of the form suggested here, such as a dark $G_2 \times SU(2)$
sector might shed light on what kind of observable features such dark
sectors might give rise to.
Note that additional dark matter such as an LSP or axions may arise
from breaking of the $E_8$, supersymmetry breaking, and the
stabilization of the moduli through the flux-generated superpotential.

\section{Distribution of flux vacua}
\label{sec:physics}

We now discuss the distribution of flux vacua and the reasons why we
believe that other F-theory vacua are strongly suppressed relative to
those on $\MM$.

\subsection{Flux vacua on $\MM$}

The counting problem of flux vacua was addressed in
\cite{Douglas-statistics, Ashok-Douglas, Denef-Douglas,
  Denef-F-theory}, building on earlier insights from
\cite{Bousso-Polchinski}.  These methods were applied 
recently in the F-theory context in
\cite{Braun-Watari1,Braun-Watari2, Watari}. Consider a fourfold $X$ with $Q=
\chi (X)/24$ and fourth Betti number $b_4\approx\chi(X)\gg Q$.  The
problem of counting vacua  can be related to the 
simplified
problem of counting the
number of lattice points in a $b_4$ dimensional sphere of radius
$\sqrt{2Q}$. In this regime, the volume of this high dimensional
sphere is not a good approximation. Instead, using the method of
\cite{Sphere,Denef-F-theory}, we can estimate the total number of
lattice points as follows:

The exact number of lattice points in this sphere  is equal to
\be
N(b_4,Q)=\frac{1}{2\pi i}\int\frac{dt}{t}e^{-Qt}Z(t),
\ee
where the contour is along the imaginary axis and passes the pole
$t=0$ on the left. (Note that with these conventions the integral runs
from $i \infty$ to $-i \infty$.)
\be
Z(t)=\sum_{\vec{n}\in\mathbb{Z}^b}e^{t\vec{n}^2/2}=\left(\sum_{n\in\mathbb{Z}}e^{tn^2/2}\right)^b
\equiv\vartheta_3(0,e^{t/2})^b,
\ee
where $\vartheta_3$ is the Jacobi theta function.

When $b$ is large, this integration can be evaluated by saddle point approximation:
\be
N(b_4,Q)\approx e^{S(t_*)},
\ee
where $t_*$ is the point where $S(t)=-\ln (-t)
-Qt+b\ln\vartheta_3(0,e^{t/2})$
takes
an extremal value.

For our case $Q\approx b/24$, $t_*=-6.18$ and
\be
N(b_4,Q)\sim 10^{3.59\times Q}
\ee
In the regime of $h^{1,1}\ll h^{3,1}$, $h^{2,1}=0$, the number of flux vacua is approximately
\be
N(h^{3,1})\sim 10^{0.9\times h^{3,1}}\label{Nflux1}
\ee

Applying this analysis to $\MM$, the total number of flux vacua is 
estimated to be
of
order $10^{272,000}$. Now, even if we only turn on the flux in
$H^4_{H*}$ which does not break any gauge symmetry, since the
dimension of $H^{2,2}_{V*}\oplus H^{2,2}_{RM*}$ is of order
$h^{1,1}\sim 250$, the total number of flux vacua is only suppressed
by a factor of $10^{50}$.

This analysis assumes among other things
that every choice of flux associated with a
set of $b$ independent integers gives rise to a solution in complex structure
moduli space associated with a  vacuum where the flux is
self-dual.  It is not clear that this is indeed the case.  To get some
sense of how this kind of condition may affect the number of vacua, we
can consider another ensemble of vacua given by the set of $G_4$ flux
that obey the self-duality relation with a fixed duality condition,
where the number of independent fluxes giving the dimension of the
sphere becomes $b_4/2$, so that we need to evaluate $N(b_4/2,Q)$.

In this case the saddle point approximation gives $t_*=-4.61$ and
\be
N(b_4/2,Q)\sim 10^{2.95\times Q},
\ee
hence approximately
\be
N'(h^{3,1})\sim 10^{0.74\times h^{3,1}}.\label{Nflux2}
\ee

Applying this formula to $\MM$, the total number of flux vacua$\sim
10^{224,000}$. While we use the formula (\ref{Nflux1}) as an estimate
for the number of flux vacua in this paper, it must be kept in mind
that this is a very rough approximation that depends on assumptions
about the solution space.  For a realistic estimate of the number of
vacua, a much more detailed analysis for this particular manifold
would need to be carried out.

\subsection{Suppression of other F-theory compactifications}

After $\MM$, the known threefold with the next largest value of
$h^{3,1}$ has $h^{1, 1}=253,h^{3,1} = 299707, \chi = 1799808$.  This
threefold is again a $B_2$ bundle over $\P^1$, with a different choice
of ``twist''\footnote{Thanks to Y.\ Huang for discussions
on this point}.  By the methods of 
estimation in the previous section, the number of flux vacua on this
threefold is suppressed by a factor of roughly ${\cal O} (10^{-3000})$
relative to those on $\MM$.  
Further known fourfolds have numbers of
flux vacua that rapidly become exponentially smaller, and the total
number of known
or inferred Calabi-Yau fourfolds is well below $10^{100}$, so known fourfolds
cannot compete with the set of flux vacua on $\MM$.  
Note that the
mirror to $\MM$ also has the same value of $\chi$, but
a smaller value of $b_4=1,214,150$, giving a number of flux vacua
smaller by a factor much smaller than ${\cal O} (10^{-3000})$.

It is natural to wonder then whether there are as yet unknown
Calabi-Yau fourfolds that can give numbers of flux vacua that dwarf
the number of vacua on $\MM$.  Since there is not even a finiteness
proof for elliptic Calabi-Yau fourfolds, this is certainly an open
question.  We briefly describe, however, some circumstantial evidence
and analogies with the threefold case that suggest that $\MM$ indeed
contributes the lion's share of 4d F-theory flux vacua and that there
are no other individual vacua that compete,
and further that there are also not
enough other elliptic fourfolds to make a significant contribution
even when all are added together.

First, we address the question of whether $\MM$ really contributes
more flux vacua than any other individual elliptic CY fourfold.  For
example, could there exist another elliptic CY fourfold with a much
larger value of $h^{3,1}$?  While we do not have any way of proving
that this does not occur, the corresponding situation for elliptic
threefolds suggests that $\MM$ really is the elliptic CY fourfold with
the largest $h^{3,1}$.  We review briefly the situation for
threefolds.  For elliptic Calabi-Yau threefolds, base surfaces can be
classified using the minimal model approach to surfaces (see, {\it
  e.g.}, \cite{bhpv}).  It was shown by Grassi \cite{Grassi} that the
only minimal surfaces that support elliptic threefolds are $\P^2$, the
Hirzebruch surfaces $\F_m, 0 \leq m \leq 12$ and the Enriques surface.
Combined with the structure of non-Higgsable clusters on base surfaces
\cite{clusters}, this gives a systematic approach to constructing all
allowed bases for elliptic Calabi-Yau threefolds \cite{mt-toric,
  Martini-WT, Johnson-WT, non-toric}.  The largest possible value of
$h^{2, 1}$ for any elliptic Calabi-Yau threefold is proven to be
$491$, for the generic elliptic threefold over $\F_{12}$ \cite{Hodge};
this threefold was constructed using toric methods in the full
analysis of Kreuzer and Skarke \cite{Kreuzer-Skarke} and is indeed the
Calabi-Yau threefold with the largest known value of $h^{2, 1}$.  The
systematic analysis of all toric bases \cite{mt-toric} and non-toric
bases supporting elliptic threefolds with $h^{2, 1}\geq 130$
\cite{non-toric} has shown that the simple toric hypersurface
construction following Batyrev \cite{Batyrev} produces a large
fraction of all possible bases that support elliptic threefolds, and
that the toric construction is particularly effective in the region of
large Hodge numbers.

\begin{figure}
\begin{center}
\includegraphics[height=10cm]{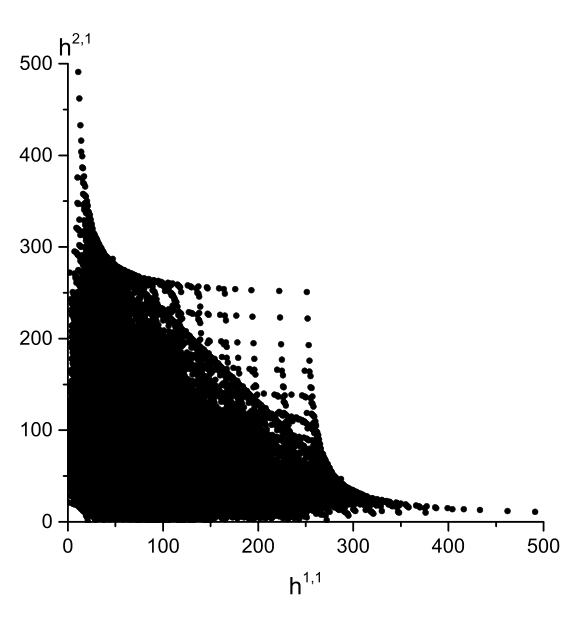}
\end{center}
\caption[x]{\footnotesize  The distribution of $h^{1, 1}$ and
  $h^{2,1}$ of the Calabi-Yau threefolds in the Kreuzer-Skarke database
 \cite{Kreuzer-Skarke}, corresponding to hypersurfaces in 4 dimensional reflexive polytopes. There is a clear ``shield'' structure and the Calabi-Yau threefold with largest $h^{2,1}$ is given by a generic elliptic fibration over $\mathbb{F}_{12}$.}
\label{f:CY3}
\end{figure}

\begin{figure}
\begin{center}
\includegraphics[height=10cm]{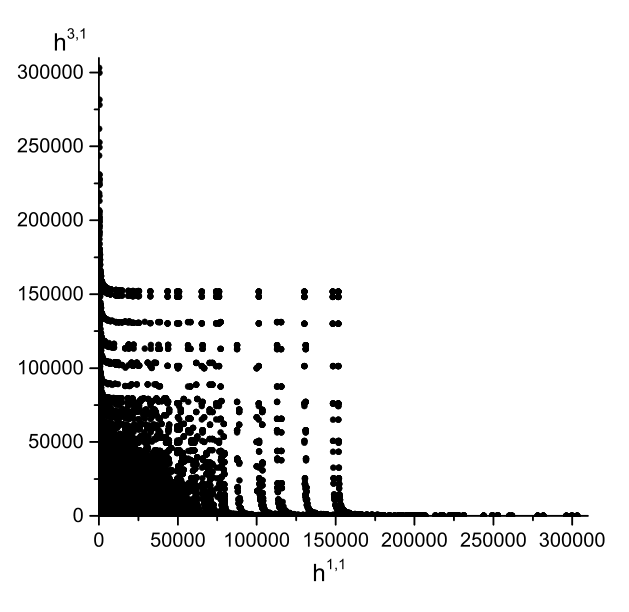}
\end{center}
\caption[x]{\footnotesize The distribution of $h^{1, 1}$ and $h^{3,1}$
  of the Calabi-Yau fourfolds in
the Kreuzer-Skarke database
  \cite{Kreuzer-Skarke-4d}, corresponding to hypersurfaces in 5
  dimensional reflexive polytopes that are weighted projective
  spaces. There is also a clear ``shield'' structure and the
  Calabi-Yau fourfold with largest $h^{3,1}$ is  $\MM$.}
\label{f:CY4}
\end{figure}

Analogizing these results from threefolds to fourfolds, it seems quite
plausible that the toric approach
is also effective in
constructing fourfolds with large $h^{3,1}$ (just as $h^{2, 1}$
parameterizes complex structure for Calabi-Yau threefolds, $h^{3,1}$
plays the same role for fourfolds).  Thus, given that several
approaches including the straightforward toric hypersurface method
have produced the fourfold with $h^{3,1} = 303,148$, which fits into
the Hodge shield of fourfolds in a parallel way to the $h^{2, 1}= 491$
Calabi-Yau threefold for threefolds (see Figure~\ref{f:CY3} and Figure~\ref{f:CY4}), it
seems plausible that there will be no elliptic fourfolds with larger
$h^{3,1}$.  Another approach to this is to consider the generalization
of the minimal model program to threefolds.  While this story involves
Mori theory \cite{Mori} and is much more complicated in various ways,
at a rough level we expect the minimal threefolds supporting elliptic
Calabi-Yau fourfolds to be one of three types: Fano threefolds, $\P^1$
bundles (or more general conic bundles) over complex surfaces $B$, and bundles of a complex surface
$B$ over $\P^1$.  Fano threefolds that have been studied in this
context have relatively small Hodge numbers
\cite{Klemm-lry}.  A systematic analysis of
$\P^1$ bundles over toric surfaces $B_2$ was carried out in
\cite{Halverson-WT}, and none of these has a value of $h^{3, 1}(X)$
that approaches 300,000.    
In the following section we describe an analysis of $B_2$ bundles over
$\P^1$ that shows that at least for toric $B_2$, no such threefold
gives a generic elliptic Calabi-Yau fourfold with larger $h^{3, 1}(X)$
than $\MM$.
Note that $\bmx$ is minimal in the class
of smooth toric threefolds in the sense that no ray can be blown down
to give another smooth toric threefold.
Any blow-up of $\bmx$ on one or more curves or  points gives a fourfold with
lower $h^{3,1}(X)$, which can be described through a singular
Weierstrass model on $\bmx$ with $(f, g)$ vanishing to orders $(4, 6)$
on curves and $(8, 12)$ at points.

Now we turn to the question of whether there are enough other elliptic
Calabi-Yau fourfolds that even if each is suppressed by a factor much
smaller than
${\cal O} (10^{-3000})$ the total number of flux vacua from all other
fourfolds might compete with the number on $\MM$.  As mentioned above,
analysis of bases for elliptic threefolds shows that the number of
non-toric bases is only larger than the number of toric bases by a
relatively small overall factor, and toric bases are particularly well
represented in the region of large $h^{2, 1}$.  In a separate work
\cite{MC} we
have recently
carried out a Monte Carlo analysis of the set of toric threefolds
that support elliptic Calabi-Yau fourfolds, including analysis of the
Hodge numbers of the corresponding fourfolds and non-Higgsable
structures on the threefolds bases.  This Monte Carlo analysis
shows that the number of toric threefold bases connected to $\P^3$ by
blow-up and blow-down transitions without encountering (4, 6) curves
is roughly
${\cal O} (10^{48})$, and analogy with the 6d case suggests that the
complete set of toric threefold bases including cases with $E_8$
non-Higgsable gauge factors and $(4, 6)$ curves is not much larger
than this, and is probably well
below ${\cal O} (10^{60})$.  This is a
large number, but nowhere near enough 
to compete with the flux vacua on $\MM$, given that the number of flux
vacua on each of these threefold bases is generally suppressed
relative to $\MM$ by a factor much smaller than ${\cal O}
(10^{-100,000}).$
Even if the number of non-toric threefold bases is much much larger
than the number of toric bases, it is hard to imagine such a massive
proliferation as would allow these bases to support fourfolds with
flux compactifications that are as numerous as those on $\MM$,
assuming as in the previous paragraph that each individual fourfold
has a value of $h^{3,1}$ that is smaller than that of $\MM$ by
thousands or tens of thousands.

We do not claim that we have proved anything here, certainly there are
many more possible ways in which elliptic Calabi-Yau fourfolds may
arise, and it is even possible that the number of such fourfolds may
be infinite.  But given the empirical information just summarized,
this seems to us to be relatively unlikely, and it seems like the
simplest situation would be that $\MM$ indeed has the largest
$h^{3,1}$ of a finite set of elliptic Calabi-Yau fourfolds that can be
constructed as fibrations over  fewer than {\it e.g.} $10^{100}$
bases, so that flux vacua on $\MM$ would dominate the set of
F-theory vacua.  Clearly, however, a more careful study
of these questions is an important issue in verifying the general
picture described in this paper.

\subsection{Other threefolds that are
$B_2$ bundles over $\P^1$}

In this section we consider $B_2$ bundles over $\P^1$, where $B_2$ is a general smooth 2d toric variety. Denote the 1d rays in the fan of $B_2$ by $\{v_i\}(1\leq i\leq N)$, then the 1d rays of a $B_2$ bundle over $\P^1$ are given by:
\be
\bsp
&w_0=(0,0,1)\\
&w_i=(v_i,0)\ ,\ (1\leq i\leq N)\\
&w_{N+1}=(m v_k+n v_{k+1},-1)\ ,\ (m,n\geq 0, 1\leq k\leq N, v_{N+1}\equiv v_1). 
\end{split}
\ee
The set of 3d cones are given by $(w_0, w_i, w_{i + 1})$
and $(w_{N+1}, w_i, w_{i + 1})$ for $1\leq i\leq (N-1)$, along with the cyclic case
$(w_0/w_{N+1}, w_{N}, w_1)$.
Hence the bundle is parametrized by a triple $(k,m,n)$. Now consider
the set of monomials in $g$, which is given by the set of lattice
points (called $A_6$ in Appendix \ref{app:tuning}) 
\be
G=\{u\in\mathbb{Z}^3|\forall w_i\ ,\ \langle u,w_i\rangle\geq -6\}.
\ee
For such a monomial $u\in G$, its  order of vanishing on a toric divisor $w_i$ is equal to
\be
\text{ord}_{w_i}(u\in G)=\langle u,w_i\rangle+6.
\ee
Its  order of vanishing on the intersection of two toric divisors $w_i$ and $w_j$ is equal to
\be
\text{ord}_{w_i w_j}(u\in G)=\langle u,w_i\rangle+\langle u,w_j\rangle+12.
\ee
Consider a monomial of the form $u=(u_2,z)$, where $u_2\in\mathbb{Z}^2$ is a 2d vector. A necessary condition for $u\in G$ is
\be
\bsp
\langle u,w_0\rangle\geq -6 &\leftrightarrow z\geq -6\\
\langle u,w_{N+1}\rangle\geq -6 &\leftrightarrow m\langle u_2,v_k\rangle+n\langle u_2,v_{k+1}\rangle-z\geq -6.
\end{split}
\ee
Hence for a given $u_2$, if there exists a monomial $u=(u_2,z)\in G$, the necessary condition is
\be
m\langle u_2,v_k\rangle+n\langle u_2,v_{k+1}\rangle\geq -12.\label{2dcond}
\ee
On the other hand, the absence of a (4,6) singularity on the intersection of $w_k$ and $w_{k+1}$ requires that there exists a monomial $u=(u_2,z)\in G$, such that
\be
\langle u_2,v_k\rangle+\langle u_2,v_{k+1}\rangle\leq -7.\label{46cond}
\ee
If $n=0$, $m\geq 13$, then from (\ref{2dcond}) we can see that
$\langle u_2,v_k\rangle \geq 0$. Then because $\langle u_2,v_{k+1}\rangle\geq -6$, the condition (\ref{46cond}) cannot be satisfied.

Following similar logic,
the only possible pairs $(m,n)$ that
satisfy these requirements are $(0,0\sim 12)$, $(1,1\sim 6)$, $(1\sim
6,1)$ and $(0\sim 12,0)$. Hence we only need to deal with a finite
number of fibrations for each $B_2$.  

The set of possible $B_2$'s is precisely the set of 61,539 toric bases
identified in \cite{mt-toric} that support 6d F-theory models. This is
because if $B_2$ has a locus with where $(f, g, \Delta)$ vanish to
orders $(4, 6, 12)$, any $B_2$ fibration over $\P_1$ also has such a
locus and it is not allowed. A subtle issue arises when there are
$(-9)-(-11)$ curves on $B_2$. In this case there are codimension two loci
(points) on these curves where $(f,g)$ vanish to orders $(4,6)$. The
solution is to blow up those points until they all become $(-12)$
curves. When a similar thing happens on a 3d base, we assume such
(non-toric) blow ups  lead to a good base in F-theory.

To compute the Hodge number $h^{3,1}$ of a generic elliptically
fibered Calabi-Yau fourfold over a 3d base, we use the following
approximate formula 
\cite{Hodge-btw}:
\begin{eqnarray}
h^{3,1} &\cong &
\tilde{h}^{3, 1}\\
&= &|F|+|G|-\sum_{\Theta\in\Delta,\dim\Theta=2}l'(\Theta)-4+\sum_{\Theta_i\in\Delta,\Theta_i^*\in\Delta^*, \dim(\Theta_i)=\dim(\Theta_i^*)=1}l'(\Theta_i)\cdot l'(\Theta_i^*)\,.\nonumber
\end{eqnarray}
Here $\Delta^*$ is the convex hull of $\{w_i\}$ and $\Delta$ is the dual polytope of $\Delta^*$, defined to be
\be
\Delta=\{u\in\mathbb{R}^3|\forall v\in\Delta^*\ ,\ \langle u,v\rangle\geq -1\}.
\ee
The symbol $\Theta$ denotes 2d faces of $\Delta$. $\Theta_i$ and
$\Theta_i^*$ denote the 1d edges of the polytopes $\Delta$ and
$\Delta^*$. $l'(\cdot)$ counts the number of integral interior points
on a face. Here $F$ is the set $A_4$ in Appendix \ref{app:tuning}, which counts the number of possible monomials in $f$.

After a thorough search among all the $B_2$ and triplets $(k,m,n)$ that
specify a $B_2$ fibration, we have
found that there is no elliptically fibered
Calabi-Yau fourfold with $h^{3,1}>303148$
over any base threefold that is a toric $B_2$ bundle over $\P^1$. In fact, all the CY
fourfolds with $h^{3,1}\gtrsim 236000$ arise from the following three
$B_2$ bases:
\be
\bsp
B_a&=(-12//-12//-12//-12//-12//-12//-11//-11,6,0)\\
B_b&=(-12//-12//-12//-12//-12//-12//-12//-11//-12,6,0)\\
B_c&=(-12//-12//-12//-12//-12//-12//-11//-12,-1,5,0)
\end{split}
\ee
The generic fibrations over $B_a$, $B_b$
 and $B_c$ give
elliptically fibered Calabi-Yau threefolds with Hodge numbers
$(h^{1,1},h^{2,1})=(222,252)$, $(251,251)$ and $(252,222)$
respectively. So they are
all located near the self-dual point $(251,251)$
on the shield (see Figure~\ref{f:CY3}). We list the largest Hodge
numbers $h^{3,1}$, that arise from
constructing threefold bases as $B_2$
bundles over $\P^1$, using these $B_2$ bases, in
Table~\ref{t:largeh31}.
(Note that $\bmx$ is one of the $B_b$ bundles over $\P^1$.)

\begin{table}
\begin{center}
\begin{tabular}{|c|c|c|}
\hline
\hline
$B_a$&$B_b$&$B_c$\\
\hline
261857 & 303148, 299707, 296266, 292825 & 261857\\
258417 & 289384, 285943, 282502, 281581 & 258416\\
254977 & 279061, 278140, 275620, 274699 & 254975\\
251537 & 272179, 271258, 268738, 267817 & 251534\\
248097 & 264376, 261856, 260935, 260014 & 248093\\
244657 & 258415, 257494, 256573, 254974 & 244652\\
243731 & 254053, 253132, 251533, 250612 & 243731\\
241217 & 249691, 248092, 247171, 246250 & 241211\\
237777 & 244651, 243730, 242809, 241210 & 237770\\
236851 & 240289, 239368, 238447, 237769 & 236849\\
 &236848&\\
 \hline
 \hline
 \end{tabular}
 \end{center}
 \caption[x]{\footnotesize List of the largest $h^{3,1}$ (greater than
   236000) of generic elliptically fibered Calabi-Yau fourfolds over a
   base in the form of a $B_{a,b,c}$ bundle over $\P_1$. }\label{t:largeh31}
 \end{table}
 
 Comparing to the largest values of $h^{3,1}$ appearing in the
 Kreuzer-Skarke database, it turns out that the values 243731, 261857,
 278140, 281581, 299707 and 303148 in \cite{Kreuzer-Skarke-4d} all
 appear in Table~\ref{t:largeh31}. Hence the known Calabi-Yau fourfolds with large $h^{3,1}$ are generally elliptically fibered. 
Since we also know that there are no non-toric base surfaces in the
same region of the Hodge plot as $B_b$ \cite{non-toric}, it seems
reasonable to conclude that no $B_2$ bundle on $\P^1$ threefold base
can give a fourfold with larger $h^{3, 1}$ then $\MM$.

\section{Possible flaws in this scenario}
\label{sec:issues}

The idea that almost all  solutions of string theory
come from compactifications of F-theory on $\MM$ depends on quite a
few assumptions, for which the level of evidence varies.  In this
section we discuss a variety of possible ways in which this scenario
may break down.  We begin with issues specific to the F-theory context
and then discuss issues that arise in a more general framework.

\vspace*{0.1in}
 \noindent
{\bf The distribution}

The Ashok-Denef-Douglas distribution of flux vacua (\ref{Nflux1}) may
be flawed or inaccurate.  The  precise distribution of flux vacua on
$\MM$ or other F-theory models should be derived more explicitly.
In particular, a more careful analysis is needed of precisely when
fluxes can give SUSY solutions where the flux is self-dual.

Even within the context of F-theory it is not clear
that all flux vacua on different bases should be weighted equally.
There are several different types of transitions between different
vacua, which may occur through tunneling processes that are wildly
different and may cause different distributions to arise dynamically.
In particular, there are Higgsing type transitions that involve
movement on the geometric moduli space, which may be lifted by the
superpotential, there are nonperturbative transitions between
different flux vacua on the same base, and there are tensionless
string type transitions between different bases.  From the point of
view of a random walk on the set of bases via blow-up and blow-down
transitions as described in \cite{MC}, the probability of reaching the
base $\bmx$ is extremely low, and also involves many transitions through
superconformal points associated with $(4, 6)$ curves.  So it is
possible that dynamics may somehow make it difficult for the theory to
reach this extreme point in the geometric moduli space of Calabi-Yau
fourfolds.  For a really sensible discussion of this kind of question,
much more powerful tools and insight are needed.

\vspace*{0.1in}
 \noindent
{\bf  Other fourfolds}

We have outlined some arguments for why we do not expect other
elliptic Calabi-Yau fourfolds with larger values of $h^{3,1}$, or
enough other elliptic fourfolds with small values of $h^{3,1}$ to
compete with flux vacua on $\MM$, but the evidence given is really
only circumstantial, and it is possible that other, as yet unknown,
elliptic Calabi-Yau fourfolds contribute more flux vacua than $\MM$.

\vspace*{0.1in}
 \noindent
{\bf  Problems on $\MM$}

Another possibility is that the set of flux vacua we have estimated on
$\MM$ may be wildly off because features specific to this geometry
make most of the vacua inconsistent for one reason or another.  For
example, a superpotential from G-flux may push most vacua to an
unphysical configuration, or one where an anthropically viable vacuum
and/or the standard model may not be realizable.  The detailed
structure of flux vacua on $\MM$ should be studied in more detail to
verify that these vacua indeed have sensible and potentially realistic
physical structure.

\vspace*{0.1in}
 \noindent
{\bf  Realizing the standard model}

For observed physics to be realized in this scenario through F-theory
on $\MM$ it seems that the standard model would need to be realized
through flux breaking of $E_8$.  The details of how this might work
would need to be worked out, perhaps along the lines of
\cite{Anderson-breaking}, to confirm that this is a possible
scenario.  Two other possible ways of realizing the standard model in the
F-theory context are through realizing part or all of the nonabelian
part of the standard model gauge group through
non-Higgsable clusters \cite{ghst}, or
through tuning  an F-theory $SU(5)$ GUT model \cite{bhv2,Donagi-Breaking}.  While the
nonabelian part of the standard model gauge group appears in F-theory
for
a wide
range of threefold bases \cite{MC}, it is not clear how or why the
$U(1)$ factor should arise in this scenario
other than possible anthropic reasons.
The $SU(3) \times SU(2)$ non-Higgsable product also does not appear on
the bases with the largest $h^{3, 1}$.  As we have discussed, it is
also not possible to tune an $SU(5)$ on the bases with large $h^{3,1}$
and many flux vacua, and even in other bases, as discussed in
\cite{Braun-Watari2} it requires tuning many moduli and is heavily suppressed
in  a natural distribution of flux vacua.  Thus, challenges remain to
see how any of the approaches to realizing the standard model will
work in detail and play a dominant role in F-theory.

\vspace*{0.1in}
 \noindent
{\bf  Other string constructions}

From our current understanding, F-theory seems to provide larger
classes of flux vacua than type IIB string theory.  Our current
understanding of flux vacua in other approaches to string
compactification is less complete; it is possible that other
constructions such as M-theory on $G_2$ manifolds, infinite families
of IIA flux compactifications with large populations at relatively
small compactification scales, non-geometric compactifications, or
other approaches may give even more numerous flux compactifications
than those of  F-theory on $\MM$.  This discussion may at least motivate a
specific search for such compactifications, and a more systematic
approach to understanding and classifying the numbers of vacua in
these other constructions.

\vspace*{0.1in}
 \noindent
{\bf    Supersymmetry and supersymmetry breaking}

In this paper we have focused on string vacuum solutions constructed
through F-theory compactifications on Calabi-Yau fourfolds.  Such
geometries are expected to be associated with theories that preserve
supersymmetry (SUSY) at some energy scale below the Planck scale.
While supersymmetry has not yet been observed in nature, it is
suggested by some aspects of observed physics (the hierarchy problem,
gauge unification, possible dark matter candidates, {\it etc.}), and
provides powerful theoretical constraints that allow us to make
analytic progress in studying large classes of string vacua.  For
supersymmetric string vacuum constructions and the results described
here to be relevant to ``real-world'' physics, either supersymmetry
must be realized at some intermediate energy scale between the TeV
scale and the Planck scale, or the qualitative results found here,
or an analogue thereof,
would have to be relevant for the class of string vacua where SUSY
is broken at the Planck/compactification scale.

Of course, it is possible that supersymmetry does not hold below the
Planck or compactification scale.  In this case new methods are needed
to analyze string compactifications without supersymmetry; for
example, F-theory would need to be expanded to describe
compactifications on elliptically fibered fourfolds without Calabi-Yau
structure.  It is possible that considerations analogous to those here
may give rise to specific geometries and/or vacuum solutions without
supersymmetry that still have specific structure analogous to the
non-Higgsable gauge groups of $\MM$, but much more work would be
needed to make any general statements of this type.

\vspace*{0.2in}

From the long list of possible issues here, it is clear that much work
remains to be done to confirm or rule out the scenario presented here
in which most solutions of string theory with low-energy SUSY would
arise from F-theory compactifications on $\MM$.  This approach does,
however, suggest a potential avenue by which string theory could
provide a predictive framework for particle physics despite the
enormous proliferation of vacua of the theory.  
While $10^{272,000}$ is a very large number of possible vacua, these
vacua will all have many qualitative similarities, and the specific
features of these vacua such as the subgroup of $G_{\rm max}$
preserved after flux breaking should be described by a calculable
statistical distribution.
Hopefully the specific
scenario we have described here is sufficiently solid and compelling
that it will stimulate further work, either to verify or to disprove
this picture of how observable physics may arise naturally in string
theory.

\vspace*{0.2in}

{\bf Acknowledgements}: We would like to thank Lara Anderson, Andreas
Braun, Will Detmold, James Gray, Jim Halverson, Yu-Chien Huang, Sam
Johnson, David Morrison, Nikhil Raghuram, Ben Safdi, Sakura Sch\"{a}fer-Nameki, Tracy Slatyer,
and Frank Wilczek
for helpful discussions.  This research was supported by the DOE under
contract \#DE-SC00012567.

\appendix
\section{Linear transformation of the polytope containing $\MM$}
\label{app:linear}

As  mentioned in \S\ref{sec:MM}, the elliptically fibered Calabi-Yau
fourfold $\MM$ can be described as a hypersurface in a 5d toric ambient space. This
toric fivefold is a $\P^{2,3,1}$ bundle over the threefold base
$\bmx$. The 3d rays $w_i(0\leq i\leq 100)$ in the toric fan of $\bmx$
are translated to the 5d rays $(w_i,-2,-3)$. In addition to that,
there are two additional vertices: $(0,0,0,1,0)$ and $(0,0,0,0,1)$,
which correspond to the $x$ and $y$ coordinates in the Weierstrass
model $y^2=x^3+fx+g$ respectively.

The convex hull of the fan of the 5d toric ambient space is a reflexive polytope $\tilde{\Delta}$ with the following 6 vertices:
\be
\bsp
&V_1=(0,0,1,-2,-3)\ ,\ V_2=(-1,-12,0,-2,-3)\ ,\ V_3=(0,1,0,-2,-3)\ ,\ \\&V_4=(0,0,0,1,0)\ ,\ V_5=(0,0,0,0,1)\ ,\ V_6=(84,492,-1,-2,-3).\label{vertices}
\end{split}
\ee

Its dual (polar) polytope (set of points $p$ satisfying $\langle p,V_i\rangle\geq -1$ for all $V_i$) $\Delta$ also has 6 vertices:
\be
\bsp
&U_1=(78,-6,3606,-1,-1)\ ,\ U_2=(35,-6,-6,-1,-1)\ ,\ U_3=(-6,1,-6,-1,-1)\ ,\ \\&U_4=(0,0,0,2,-1)\ ,\ U_5=(0,0,0,-1,1)\ ,\ U_6=(78,-6,-6,-1,-1).\label{dvertices}
\end{split}
\ee

The toric fivefold in \cite{Candelas-Skarke} which contains the Calabi-Yau fourfold $\MM$ is a weighted projective space $\P^{1,84,516,1204,1806}$. It is a (singular) toric variety with 6 vertices in the fan:
\be
\bsp
&V_1'=(1,0,0,0,0)\ ,\ V_2'=(0,1,0,0,0)\ ,\ V_3'=(0,0,1,0,0)\ ,\ \\&V_4'=(0,0,0,1,0)\ ,\ V_5'=(0,0,0,0,1)\ ,\ V_6'=(-1,-84,-516,-1204,-1806).
\end{split}
\ee

There exists an SL(5) transformation matrix that transforms the $V_i$ into $V_i'$:
\be
\begin{pmatrix}
0&0&1&0&0\\
-1&0&0&0&0\\
-12&1&0&0&0\\
-26&2&2&1&0\\
-39&3&3&0&1
\end{pmatrix}
\cdot
\begin{pmatrix}
0&-1&0&0&0&84\\
0&-12&1&0&0&492\\
1&0&0&0&0&-1\\
-2&-2&-2&1&0&-2\\
-3&-3&-3&0&1&-3\\
\end{pmatrix}
=
\begin{pmatrix}
1&0&0&0&0&-1\\
0&1&0&0&0&-84\\
0&0&1&0&0&-516\\
0&0&0&1&0&-1204\\
0&0&0&0&1&-1806\\
\end{pmatrix}.
\ee

\section{The
Weierstrass model on $\bmx$ and the possibility of tuning}
\label{app:tuning}

The set of monomials in the line bundle $\mc{O}(-nK)$ on $\bmx$ is given by the set 
\be
A_n=\{u\in\mathbb{Z}^3|\forall w_i\ ,\ \langle u,w_i\rangle\geq -n\}.\label{monomialset}
\ee
For such a monomial $u\in A_n$, its  order of vanishing on a toric divisor $w_i$ is equal to
\be
\text{ord}_{w_i}(u\in A_n)=\langle u,w_i\rangle+n.
\ee

A general Weierstrass model on $\bmx$ can be written in the following Tate form:
\be
y^2+a_1 xy+a_3 y=x^3+a_2 x^2+a_4 x+a_6\label{Tateform}
\ee
The coefficients $a_1\in\mc{O}(-K)$, $a_2\in\mc{O}(-2K)$, $a_3\in\mc{O}(-3K)$, $a_4\in\mc{O}(-4K)$, $a_6\in\mc{O}(-6K)$. The  order of vanishing of $a_n$ on a divisor $w_i$ is equal to
\be
\text{ord}_{w_i}(a_n)=\min(\langle u,w_i\rangle+n)|_{u\in A_n}.
\ee

The relations between these Tate coefficients and the functions $f$ and $g$ in the Weierstrass form:
\be
y^2=x^3+f x+g
\ee
are \cite{Bershadsky-all}
\be
\bsp
f&=\frac{1}{48}(-(a_1^2+4a_2)^2+24a_1 a_3+48 a_4)\\
g&=\frac{1}{864}(-(a_1^2+4a_2)^2+36(a_1 a_3+2a_4)(a_1^2+4a_2)-216(a_3^2+4a_6)).
\end{split}
\ee 
For a general elliptically fibered Calabi-Yau
describe through Weierstrass model, 
the set of monomials in $f$ and $g$ are simply given by the sets $A_4$
and $A_6$ in (\ref{monomialset}).

The correspondence between the order of vanishing of those
coefficients on a divisor $D$ and the gauge group on $D$ can be found
in \cite{Bershadsky-all}. For example, if we want to tune an SU(5)
gauge group on divisor $w_i$, then the order of vanishing of
$(a_1,a_2,a_3,a_4,a_6)$ on $w_i$ should be $(0,1,2,3,5)$. This is
impossible for a generic fibration on any base. However, we may remove
a minimal number of points in $A_n$ (or equivalently, tune the
coefficients of these monomials to be zero), so that this condition is
satisfied. Since the order of vanishing will always increase after
this tuning process, the value $\text{ord}_{w_i}(a_1)$ should be
equal to
0 before the tuning. The only divisors satisfying this constraint are
\be
w_0=(0,0,1)\ ,\ w_1=(-1,-12)\ ,\ w_2=(0,1)\ ,\ w_{100}=(84,492,-1).
\ee However, after explicit computation, the tuning on these divisors
always gives rise to codimension-1 locus with
$(\text{ord}(f),\text{ord}(g))=(4,6)$. This codimension-1 singularity
is non-minimal and not allowed in F-theory. Hence we cannot tune an
SU(5) gauge group on $\bmx$.

A similar analysis can be applied for other gauge groups. It turns out
that even the slightest enhancement of the Tate form ({\it e.g.}  to SU(2) on some
divisor) is not allowed on $\bmx$, due to the fact that $\bmx$ is a
$B_2$ bundle over $\P^1$, and the base $B_2$ contains a lot of -12,
-1, -2, -2, -3, -1, -5, -1, -3, -2, -2, -1 chains with ``saturated''
gauge groups.

\end{document}